\documentclass[%
 reprint,
groupedaddress,
superscriptaddress,
 amsmath,amssymb,
 aps,
 prl,
 showpacs
floatfix,
]{revtex4-2}
\usepackage{graphicx}
\usepackage{alphabeta}
\usepackage{makecell,booktabs}
\usepackage{dcolumn}
\usepackage{bm}
\usepackage{hyperref}
\hypersetup{
    colorlinks=true,
    urlcolor= blue,
    citecolor=blue,
linkcolor= blue}
\usepackage{xcolor}

\renewcommand{\vec}[1]{{\textbf{\textit{#1}}}}

\newcommand{\Rxx}{$R_{xx}$}
\newcommand{\Rxy}{$R_{xy}$}
\newcommand{\Rxxsym}{$R_{xx,{{\text{sym}}}}$}

\newcommand{\Rxysym}{$R_{xy,{{\text{sym}}}}$}
\newcommand{\Rxyasym}{$R_{xy,{{\text{asym}}}}$}

\newcommand{\Tco}{$T_{c,{{\text{onset}}}}$}
\newcommand{\Tcoff}{$T_{c,{{\text{offset}}}}$}
\newcommand{\Tc}{$T_c$}
\newcommand{\Rn}{$R_N$}

\begin{document}


\title{Emergent anisotropic three-phase order in critically doped superconducting diamond films}

\author{Jyotirmay Dwivedi}
\author{Jake Morris}
\author{Saurav Islam}
\affiliation{Department of Physics, The Pennsylvania State University, University Park, PA 16802 USA}
\author{Kalana D. Halanayake}
\author{Gabriel A. Vázquez-Lizardi}
\affiliation{Department of Chemistry, The Pennsylvania State University, University Park, PA 16802 USA}
\author{David Snyder}
\affiliation{Applied Research Lab, The Pennsylvania State University, University Park, PA 16802 USA}
\author{Anthony Richardella}
\affiliation{Department of Physics, The Pennsylvania State University, University Park, PA 16802 USA}
\affiliation{Materials Research Institute, The Pennsylvania State University, University Park, PA 16802 USA}
\author{Luke Lyle}
\affiliation{Applied Research Lab, The Pennsylvania State University, University Park, PA 16802 USA}
\author{Danielle Reifsnyder Hickey}
\affiliation{Department of Chemistry, The Pennsylvania State University, University Park, PA 16802 USA}
\author{Nazar Delegan}
\author{F. Joseph Heremans}
\affiliation{Q-NEXT, Argonne National Laboratory, 
Chicago, IL 60439 USA}
\author{David D. Awschalom}
\affiliation{Q-NEXT, Argonne National Laboratory, 
Chicago, IL 60439 USA}
\affiliation{Pritzker School of Molecular Engineering, University of Chicago, Chicago, IL 60637 USA}
\author{Nitin Samarth}
\affiliation{Department of Physics, The Pennsylvania State University, University Park, PA 16802 USA}
\affiliation{Materials Research Institute, The Pennsylvania State University, University Park, PA 16802 USA}
\affiliation{Q-NEXT, Argonne National Laboratory, 
Chicago, IL 60439 USA}
\affiliation{Department of Materials Science and Engineering, The Pennsylvania State University, University Park, PA 16802 USA}

\email{nxs16@psu.edu}

\begin{abstract}
Two decades since its discovery, superconducting heavily boron-doped diamond (HBDD) still presents unresolved fundamental questions whose resolution is relevant to the development of this material for quantum technologies. We use electrical magnetotransport measurements of critically-doped homoepitaxial single crystal HBDD films to reveal signatures of intrinsic (electronic) granular superconductivity. By studying the dependence of electrical resistivity on temperature and magnetic field vector, we infer that this granularity arises from electron correlations. This is revealed by a striking three-phase anisotropy in the magnetoresistance, accompanied by a spontaneous transverse voltage (Hall anomaly). Our findings indicate an emergent magnetically tunable intrinsic order in an otherwise isotropic three dimensional single crystal HBDD film, offering new insights into the mechanism of superconductivity in this quantum material.
\end{abstract}

\maketitle

Superconducting heavily boron-doped diamond (HBDD) is a promising material for `quantum-on-chip' architectures, with the potential to seamlessly bridge superconducting qubits such as transmons and single spin quantum defects such as nitrogen vacancy (NV) centers~\cite{Marcos_PhysRevLett.105.210501, Zhu_Nat.478.7368}. Initially discovered in polycrystalline diamond synthesized using high-pressure-high-temperature techniques~\cite{Ekimov_Nature.428.542}, superconductivity has also been observed in polycrystalline~\cite{Takano_APL,Winzer_PhysicaC.432.65, Zhang_PhysRevB.84.214517} and homoepitaxial~\cite{Yokoya_Nat2005.438.7068,Takano_JPhysCondensMatter.21.253201,Okazaki_APL.106.052601,Lin_AdvFuncMat} HBDD thin films synthesized via microwave plasma chemical vapor deposition (MPCVD). Controlling the superconducting behavior of HBDD is important in the context of quantum technologies, thus providing a strong motivation for understanding its origins. Yet, despite extensive research since its discovery, a rigorous understanding of superconductivity in HBDD is still being sought~\cite{Blase_NatMater.8.375,Bustarret_PhysicaC}. For example, although a consensus has emerged that key characteristics of superconductivity in HBDD are consistent with a weak coupling BCS superconductor in the dirty limit~\cite{Bustarret_PhysicaC}, the potential role of electron correlations remains an unsettled question \cite{Blase_NatMater.8.375}. Given tantalizing experimental advances such as the observation of a superconducting transition temperature (\Tc) as high as 25 K in homoepitaxial HBDD films grown on (111) diamond substrates \cite{Okazaki_APL.106.052601} and theoretical predictions of even higher \Tc~\cite{Shirakawa_JPSJ.76.014711,Sakai_PhysRevMaterials.4.054801}, continued exploration of experimental insights into the pairing mechanism remains important.

In this Letter, we report the observation of emergent electronic granularity accompanied by striking magnetic field- and temperature-dependent spatial anisotropy in the magnetoresistance (MR) and Hall effect in homoepitaxial single crystal HBDD films across a Mott metal-superconductor phase transition. We attribute this behavior to electron correlations in a doping regime close to the metal-insulator transition, determined by a critical boron doping density $n_{Bc} \sim 4 \times 10^{20}$ cm$^{-3}$~\cite{Bustarret_PhilTrans,Klein_PhysRevB.75.165313}.     
An ``impurity band resonating valence bond'' (IBRVB) model of Cooper pairing has raised the possibility that correlations may play a key role in superconductivity in HBDD ~\cite{Baskaran_2006,Baskaran_JSNM.21.45}. 
When the boron doping exceeds $n_{Bc}$, delocalized acceptor states that form an impurity band create a disordered Mott metal in the normal state~\cite{Inushima_PhysRevB.79.045210, Masayuki_DiamondRelatMater.20.1357, Deneuville_DiamondRelatMater.07.915, Kumar_JPhysCommun.2.045015}. At low temperatures, strong electron correlations between neighboring neutral boron acceptors (B\textsuperscript{0}–B\textsuperscript{0}) form superconducting bosonic channels. If the boron concentration is near the Mott transition, small fractions of B\textsuperscript{+} and B\textsuperscript{-} free carriers can spontaneously delocalize, simultaneously creating fermionic channels. The competition between these bosonic and fermionic channels can then lead to inhomogeneous (granular) superconductivity. This  electronic (or intrinsic) granularity \cite{Kapitulnik_RevModPhys.91.011002, Sacepe_NatPhys.16.734} is fundamentally different from structural (or extrinsic) granularity that arises from weak-linked grain boundaries in polycrystalline or layered materials  \cite{Aguilar_JSNM.36.1835, Klemencic_Carbon.175.43,Zhang_ACS.Nano.11.11746,Charikova_PMM.124.670, Bustarret_PTRSA.366.267}. The intrinsic granularity in the IBVRB model stems from disorder-induced Coulomb repulsion or scattering, akin to the physics underlying the superconductor-insulator transition in other disordered superconductors \cite{Sacepe_NatPhys.7.239, Chand_PRB.85.014508}. An intrinsic granular superconductor can be visualized as a network of superconducting islands embedded in a metallic/insulating matrix (Fig.~\ref{Fig EM3})~\cite{Kapitulnik_RevModPhys.91.011002, Sacepe_NatPhys.16.734}. Our transport measurements are indeed consistent with such a picture, leading us to propose that experimental evidence of electronic granularity in HBDD offers evidence of electron correlations in the pairing mechanism. 

Signatures of granular superconductivity have been reported in prior studies of nano- and poly-crystalline HBDD films. However, in such samples, it is difficult to disentangle intrinsic origins of this granularity from the extrinsic origins due to microstructure (e.g. superconducting grains weakly linked via amorphous grain boundaries) ~\cite{Klemencic_Carbon.175.43,Zhang_PhysRevApplied.6.064011, Zhang_PhysRevApplied.12.064042, Sobaszek_Carbon.119337, Zhang_ACS.Nano.11.11746}.
Intragrain intrinsic granularity has also been observed in nano- and poly-crystalline HBDD where it is attributed to preferential boron incorporation and doping-induced disorder \cite{Willems_PhysicaC.470.S588, Zhang_AdvMater.26.2034}. Controlling this disorder could lead to significantly higher $T_c$ than realized in HBDD thus far, approaching that of MgB\textsubscript{2} (39 K) \cite{Shirakawa_JPSJ.76.014711}. Motivated by these still unresolved questions, we report a systematic set of electrical MR measurements of homoepitaxial HBDD films of high structural quality, wherein sources of extrinsic granularity are minimized. This provides an attractive route to explore intrinsic granularity in this disordered superconductor and allows us to more clearly examine the effect of electron correlations on the superconducting state.

We synthesize HBDD films on single crystal (001) electronic and optical grade diamond substrates via MPCVD using BCl$_3$ as a dopant precursor gas; this halide growth chemistry contrasts with the more widely used hydrocarbon-based growth achieved using di-borane/tri-methyl borane. A series of samples with thicknesses in the range 0.5-20 \textmu m with different boron concentrations are investigated (see the Supplementary Materials for more details~\cite{Supplementary_Materials}); here we focus on a 0.5 \textmu m thick film (AE-1) grown on an electronic grade (001) diamond substrate. As shown in the Supplementary Materials ~\cite{Supplementary_Materials}, the findings reported here are reproduced in other samples. Detailed structural characterization using spatial Raman spectroscopy, atomic force microscopy (AFM), and transmission electron microscopy (TEM) shows uniform doping and single crystal growth over length scales of greater than 9 \textmu m \cite{Supplementary_Materials}. The sample has an estimated boron concentration of $5 \pm 0.4 \times 10^{20}$ cm\textsuperscript{-3} (Fig.~\ref{Fig EM1}). We note that, as found in past studies of HBDD~\cite{Klein_PhysRevB.75.165313}, the carrier density determined by the Hall effect is much higher than might be expected from the boron doping (see Supplementary Materials~\cite{Supplementary_Materials}). The boron concentration and the temperature variation of the conductivity (Fig. S3(a) \cite{Supplementary_Materials})places our sample on the metallic side of the critical regime of a Mott insulator-to-metal transition, making it viable to probe the anomalous transport behavior due to competing electron correlation and electron-phonon coupling \cite{Okazaki_APL.106.052601, Winzer_PhysicaC.432.65, Baskaran_JSNM.21.45, Klein_PhysRevB.75.165313, Bustarret_PTRSA.366.267}.

\begin{figure*}
    \centering
    \includegraphics[width=\textwidth]{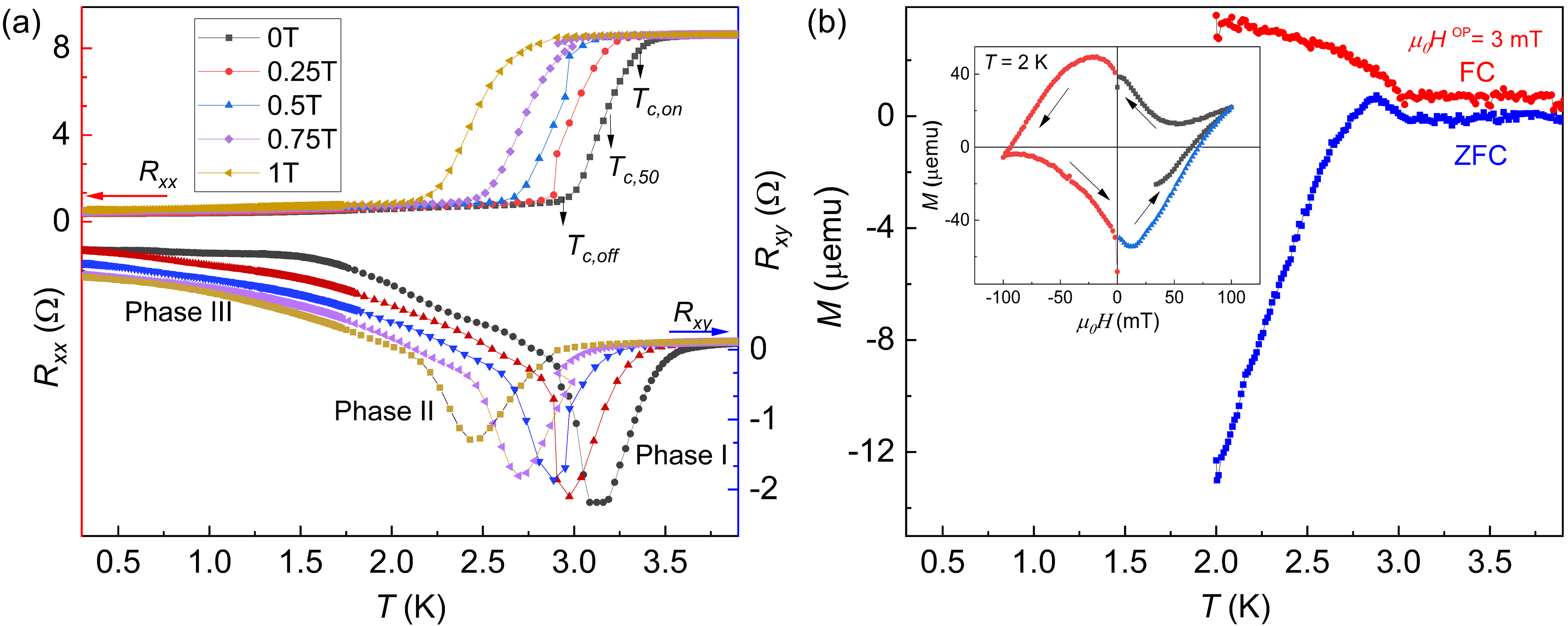}
    \caption{(a) Temperature dependence of \Rxx~(left axis, upper dataset) and \Rxy~(right axis, lower dataset) at various values of $H$ applied perpendicular to the sample plane. (b) Temperature dependence of the magnetization, $M$, showing field cooled (FC) and zero field cooled (ZFC) curves separating at around $T = 2.8$~ K. Inset shows the hysteretic magnetic field dependence of $M$ in the superconducting state at $T = 2$~K due to Meissner screening.}\label{Fig 1}
\end{figure*}

We measure electrical transport as a function of temperature ($T$) and magnetic field ($H$) using a measurement geometry shown in the End Matter (Fig.~\ref{Fig_EM3}). Figure \ref{Fig 1}(a) shows the temperature variation of the longitudinal resistance, \Rxx,  and the transverse or Hall resistance, \Rxy, at different values of $H$ applied out-of-plane (OP) to the film. The \Rxx$(T)$ data show the onset of superconductivity at \Tco $\approx$ 3.3 K at zero magnetic field. (\Tco~ is defined as the temperature where the \Rxx~drops to 90\% of the normal state value \Rn~at $T=4$~K.) With increasing $H$, \Tco~decreases as expected. The temperature variation of \Rxy, measured simultaneously with \Rxx, shows a spontaneous Hall voltage emerge below \Tco, even at $H = 0$. With decreasing temperature, the Hall voltage goes through a minimum at around $T = T_c^{50\%}$, where $R_{xx}=0.5 R_N$, before reversing sign and then saturating at $T \leq$ \Tcoff; this phenomenon is known as a ``Hall anomaly.'' The origins of the Hall anomaly have been debated extensively for other superconductors \cite{Vasek_PhysicaC.411.164, Hagen_PRB.43.6246, Bozovic_CM.8.7}, with explanations that include order parameter confinement and vortex flow in high temperature cuprates \cite{Zechner_PhysicaC.533.144}, electronic nematicity in graphene \cite{Cao_Science.372.264}, and granularity in Pb films \cite{Segal_PRB.83.094531}. A Hall anomaly has been reported in HBDD in a couple of prior papers and explained by structural granularity \cite{Sobaszek_Carbon.119337,Bhattacharyya_arXiv.1710.05170}. However, the presence of a Hall anomaly in our single crystal samples without any structural granularity points to intrinsic origins that we will discuss later in the text.

To further confirm the superconducting transition, we used magnetometry to probe the diamagnetism in our HBDD films at $T \leq$ \Tcoff. The temperature and magnetic field dependent magnetization ($M$) of an HBDD film is shown in Fig.~\ref{Fig 1}(b). The variation of $M$ with decreasing temperature shows the separation of the zero field cooled (ZFC) and field cooled (FC) magnetic response at around $T = 2.8$ K, marking the onset of diamagnetism. In the case of complete Meissner screening, the FC curve should saturate at $T =$ \Tcoff which should be approximately equal to the temperature at which the resistance vanishes. In our case, at the minimum accessible temperature ($T \approx 2$ K), the ZFC signal does not saturate, while the resistance saturates by $T = 2$ K. This disparity between the onset temperatures of low resistance and diamagnetism is yet another indictor of granularity. In polycrystalline HBDD, zero-resistance percolation paths form across superconducting grains before global coherence emerges, leading to a drop in resistance prior to complete Meissner screening~\cite{Aguilar_JSNM.36.1835, Klemencic_Carbon.175.43, Zhang_ACS.Nano.11.11746, Charikova_PMM.124.670, Bednorz_EurophysicsLetters.3.379, Aguilar_JSNM.36.1835, Yonezawa_PhysRevB.97.014521, Zhang_AdvMater.26.2034, Kumar_JPhysCommun.2.045015}. However, its occurrence in homoepitaxial single-crystal samples points to intrinsic granularity. The magnetization versus magnetic field dependence (inset Fig.~\ref{Fig 1}(b)) shows diamagnetism at $T = 2$ K. Note that all magnetization data are presented after subtracting the paramagnetic and diamagnetic background signals from the substrate and sample holder (Fig. S3(b),(c) in the Supplementary Material~\cite{Supplementary_Materials}). The slight asymmetry in the $M-H$ loop is the result of an inhomogeneous screening current distribution, reaffirming the inhomogeneous superconductivity.

Our first key observation is the saturation of \Rxx~to a finite resistance of $\sim 0.1 R_N$ at $T \leq$\Tcoff$\approx 2.8$ K (Fig.~\ref{Fig 1}(a)). Similar anomalous transport in polycrystalline HBDD has been explained by modeling the system as a disordered array of Josephson junctions~\cite{Zhang_ACS.Nano.11.11746,Zhang_PhysRevApplied.6.064011}. The temperature variation of \Rxx~is equivalent to that of a parallel resistor network with competing low resistance bosonic channels and high resistance fermionic channels. Our second key observation, the Hall anomaly, can also be explained with a similar resistor network model. Superconductors such as MgB\textsubscript{2} and twisted bilayer graphene show this effect, with different possible origins like vortex-anti vortex unbinding~\cite{Vasek_PhysicaC.411.164} or nematicity~\cite{Cao_Science.372.264,Bozovic_CM.8.7}. Despite the differences in material systems, the magnitude of the Hall anomaly is always proportional to d\Rxx/dT~\cite{Segal_PRB.83.094531}. These two phenomenological models based on granular superconductivity neatly fit our data, as shown in Fig.~\ref{Fig EM2}~ \cite{Supplementary_Materials}.The parameters obtained are comparable to those obtained in the literature for polycrystalline samples~ \cite{Zhang_ACS.Nano.11.11746,Zhang_PhysRevApplied.6.064011}. These models help us identify a three step transition from metallic to an inhomogeneous superconducting phase as shown by Phase I, Phase II and Phase III in Fig.~\ref{Fig 1}(a). This inhomogeneity might arise due to preferential boron incorporation in our samples but structural characterization suggests uniform doping~\cite{Supplementary_Materials}. It is more likely that the proximity to the critical disorder of insulator-metal transition is the main cause of granularity, where strong electron correlations and electron-phonon coupling compete with each other in the superconducting phase \cite{Klein_PhysRevB.75.165313, Bustarret_PTRSA.366.267}.

\begin{figure*}
    \centering
    \includegraphics[width=\textwidth]{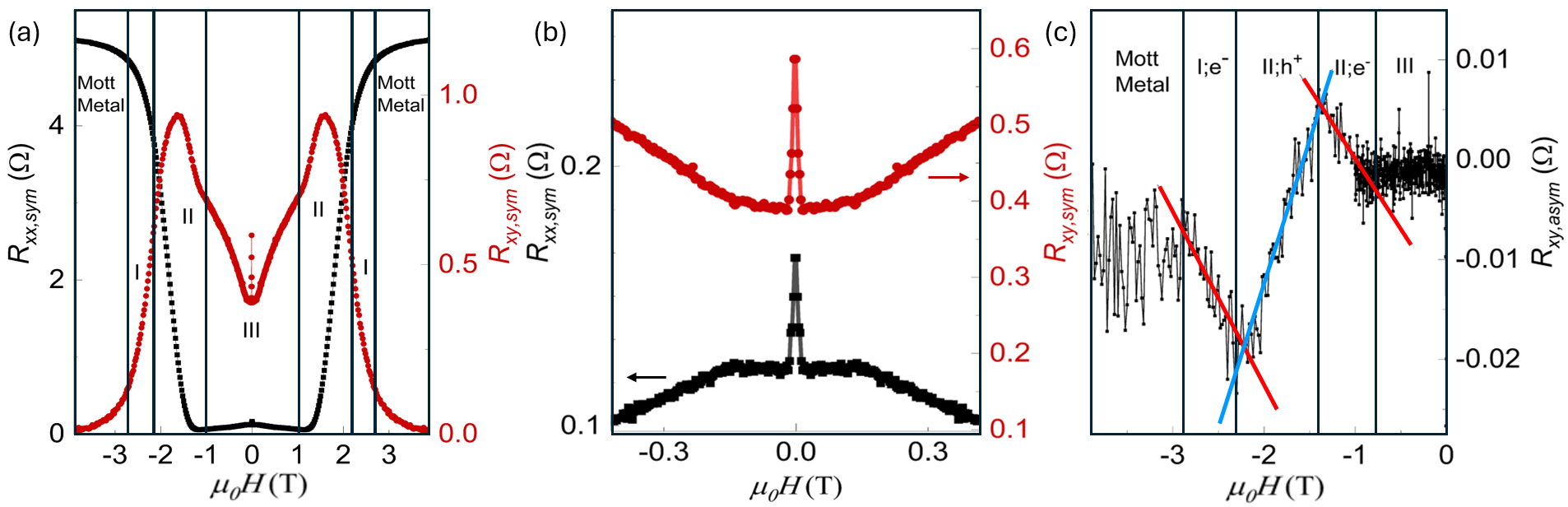}
   \caption{(a) Left axis (black) shows the variation of \Rxx~with applied OP field at $T = 2$~K. Right axis (red) is the symmetric component of transverse resistance (ETV). Regions with different transport phases are marked as I, II and III. Red regions represent fermions and blue regions represent bosons. (b) Expanded view of boson dominated Phase III highlighting the the peak feature around zero field (c) Anti-symmetric (Lorentz force) component of \Rxy~showing different Hall coefficients in regions I, II and III corresponding to (a).
}\label{Fig 2}
\end{figure*}

The three phase granularity is further reinforced by the longitudinal MR (\Rxx($H$)) and transverse MR (\Rxy($H$)) data (Fig.~\ref{Fig 2}). \Rxx~increases with increasing magnetic field when the OP $H$ exceeds $H_{c1}$, with normal state resistance restored above $\mu_0 H_{c2} \approx$~3 T (Phase I). In the superconducting state ($0 \leq H \leq H_{c1}$), the residual \Rxx~decreases with increasing field at different rates in Phases II and III, with a sharp peak around zero field (Fig.~\ref{Fig 2}(b)). The peak near $H=0$ and negative low-field MR below $H_{c1}$ has been interpreted as a signature of weak localization resulting from strong electron-electron interaction in the fermionic channels ~\cite{Blase_NatMater.8.375,Bhattacharyya_NJP.22.093039,Sacepe_NatPhys.7.239,Klein_PhysRevB.75.165313, Bustarret_PTRSA.366.267}.
In the normal state, \Rxy~ is antisymmetric with respect to the OP magnetic field because of the Lorentz force acting on the charge carriers and shows typical hole transport for HBDD~\cite{Winzer_PhysicaC.432.65}. Within the superconducting transition, \Rxy~becomes symmetric in field, manifesting as an ``even-in-field transverse voltage'' (ETV)~\cite{Zhao_Nanomaterials.12.1313,Xu_NatCommun.13.5321,Jin_PRB.64.220506,Segal_PRB.83.094531}. After correcting for contact misalignment and separating the antisymmetric Lorentz force components ~\cite{Supplementary_Materials}, the ETV (Fig.~\ref{Fig 2}(a), red curve) is found to be an order of magnitude larger than the conventional Hall effect. It also exhibits a peak near zero field (Fig.~\ref{Fig 2}(b)), mirroring the behavior of \Rxx~and supporting the presence of weak localization in residual fermionic channels~\cite{Blase_NatMater.8.375,Bhattacharyya_NJP.22.093039,Sacepe_NatPhys.7.239, Bousquet_PhysRevB.95.161301, Klemencic_Carbon.175.43, Klein_PhysRevB.75.165313}. The Lorentz force component (Hall effect) of the transverse MR (Fig.~\ref{Fig 2}(c)) shows that the carrier type switches from electron to hole to electron across the three phases of granular superconducting transition with reducing carrier concentration, suggesting that the fermionic channels ``shrink" with temperature. This switching of charge carrier type can be explained using the IBRVB model where B\textsuperscript{+} and B\textsuperscript{-} species are spontaneously created in the impurity band~\cite{Baskaran_JSNM.21.45}. Different carrier properties extracted from Fig.~\ref{Fig 2}(c) are summarized in Table II of the Supplementary Material~\cite{Supplementary_Materials}. The varying MR slopes in both \Rxx~and \Rxy~in Phases I, II, and III in Fig.~\ref{Fig 2} reflect transitions between a metallic phase, competing fermionic-bosonic phase, and boson-dominated transport, respectively.

\begin{figure}
    \centering
    \includegraphics{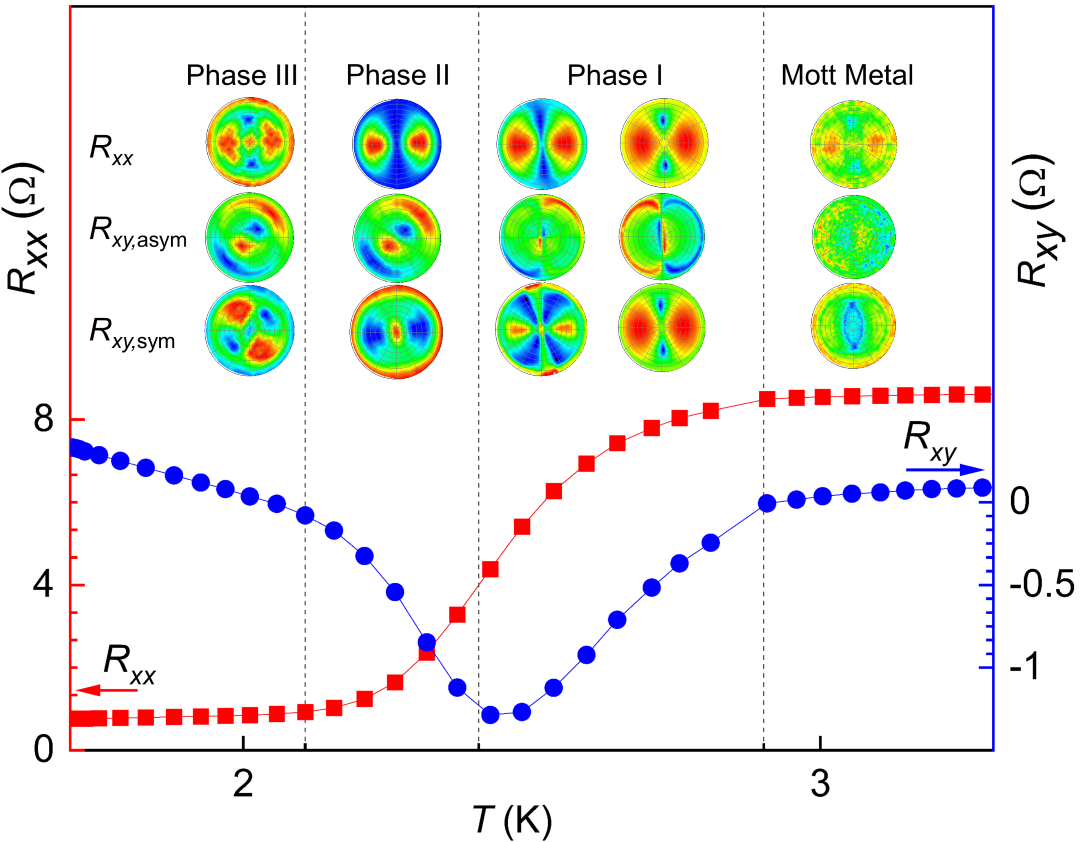}
       \caption{Evolution of anisotropy symmetry and magnitude in the MR across different resistive phases demarcated by dotted lines.
}
    \label{Fig 3}
\end{figure}

To determine whether this granularity is intrinsic or extrinsic, we measure the magnetic field orientation dependence of \Rxx~and \Rxy~at fixed temperatures below \Tco. This is done with a fixed magnetic field magnitude while rotating its orientation in 5$^{\circ}$ steps over the unit sphere. The value of \Rxx(\Rxy) recorded at each field vector is symmetrized (antisymmetrized) with respect to field direction to obtain \Rxxsym(\Rxyasym)~\cite{Supplementary_Materials}. These corrections account for any inevitable sample misalignment. We plot this data as azimuthal equidistant projections, summarized in Fig.~\ref{Fig 3} (See End Matter, Fig.~\ref{Fig EM4} for more details). High resistance (red) regions lying in a 45$^{\circ}$ conical section around the equatorial plane correspond to higher IP magnetic field components. Low resistance (blue) regions are confined around the poles and correspond to the OP field components. The color intensity is proportional to the magnitude of the angular MR variation within each plot. For purely 3D isotropic superconductors, no change is expected in resistance as the magnetic field angle is changed. For 2D superconductors, an OP-to-IP field rotation results in a decrease in resistance because electron pairing is more robust against magnetic field in the IP direction~\cite{Shiogai_PRB.97.174520}. Here, we see the {\it opposite} effect where OP-to-IP $\vec{H}$ rotation increases resistance when $\vec{H}  \perp \vec{J}$. A similar observation was reported in 3D polycrystalline HBDD samples that have vertical grain boundaries~\cite{Zhang_PhysRevApplied.12.064042}. However, our HBDD film is a 0.5 \textmu m thick 3D single crystal with no evident grain boundary formation; yet, the granular superconductivity models apply well to our data. This indicates that the anomalous MR anisotropy in HBDD emerges due to intrinsic granularity and critical doping concentration.

Tracking the evolution of this anisotropic angular MR behavior at different $H-T$ points can provide further insight into the nature of the three-phase superconducting transition identified previously. We plot the equidistant azimuthal projections of the angular MR at $T =$ 300 mK, 2.1 K, 2.6 K, 2.9 K, and 4 K with different magnitude of magnetic field ($\mu_0 H =$ 0.25 T, 0.5 T, 0.75 T, and 1 T). Then, we extract \Rxxsym, \Rxyasym~and \Rxysym~from each, resulting in a dataset of more than 50 plots (Figs. S6-S8 in the Supplementary Material~\cite{Supplementary_Materials}). Careful inspection indicates that the angle-dependent MR anisotropy shows similar symmetry and magnitude for all $H-T$ points lying within a single phase. The anisotropy varies as we move across the three phases. This is visualized in Fig.~\ref{Fig 3} where the top, middle, and bottom panels represent \Rxxsym, \Rxyasym, and \Rxysym, respectively. The reference blue and red curves are extracted from Fig.~\ref{Fig 1}(a) and represent the variation of \Rxx~and \Rxy~with temperature for $\mu_0 H_{\textsubscript{OP}} = 1$ T. 

The normal metallic state at $T =4$~K shows negligible variation in resistance with change in field angle, restricting the three phase anisotropy to the superconducting transition. In Phase I, at \Tco~= 2.9 K , the angular MR signal shows a two-fold symmetry for both \Rxxsym~and \Rxyasym~with two high resistance (red) lobes for $\vec{H}_{xy}  \perp \vec{J}$ and two low resistance (blue) pockets for $\vec{H}_{xy}  \parallel \vec{J}$. The large MR in phase I implies that most of the transport in this phase is fermionic with few bosonic channels operating. At the boundary of phase I and phase II, near $T = 2.62$ K, \Rxx~drops below $0.5 R$\textsubscript{N} and \Rxy~is at a minimum. Here, the bosonic channels start to compete with fermionic channels leading to enlarged low resistance zones in \Rxxsym~with \Rxysym~showing a different 4-lobed symmetry. Within phase II, at 2.1 K, bosonic channels start to dominate: this is indicated by a further expansion of the low resistance blue regions in \Rxxsym, although with the same two-fold symmetry as phase I. In contrast, \Rxysym~shows a complimentary symmetry with blue lobes aligned with $\vec{H}_{xy}  \perp \vec{J}$ and red lobes aligned with $\vec{H}_{xy}  \parallel \vec{J}$. This symmetry flip coincides with the fact that $dR_{xy}/dT$ and $dR_{xx}/dT$ are approximately 90$ ^{\circ}$~relative to each other in this phase, reinforcing Segal's model~\cite{Segal_PRB.83.094531} used to fit the Hall anomaly data in Fig.~\ref{Fig EM2}. Measurements for phase III are conducted at $T = 300$~mK where the transport is overtaken by the bosonic channels with a few residual fermionic channels. This reduces the magnitude of the angular MR for both \Rxxsym~and \Rxysym, as shown by a lower contrast in the polar plots. Two-fold symmetry still persists in \Rxxsym~accompanied by 45$^{\circ}$ rotation about the polar axis for \Rxysym, corresponding to the is 45$^{\circ}$ relative slope between $dR_{xy}/dT$ and $dR_{xx}/dT$. The angular MR in \Rxyasym~(middle panel) is an order of magnitude lower than that of \Rxysym~and is measurable only when $\vec{H}$ is OP. This results in the blue/red pockets emerging only near the poles of the projection plots. This is expected because \Rxyasym~is a consequence of the Lorentz force which is non-zero only for an OP field component. 
The variation of magnitude in angular MR across the three phases of the superconducting transition reflects the competition between low-resistance bosonic channels and high resistance fermionic channels, establishing the tunable granularity of superconductivity in these HBDD films. We observe similar behavior in other samples as well (Fig. S10-S11 in the Supplementary Material~\cite{Supplementary_Materials}). The simultaneous evolution of angular MR symmetries reveals the emergence of an anisotropic order within this disordered granular superconductor which is dependent on $\vec{J} \cdot \vec{H}$ and temperature.

Our measurements indicate that the granular superconductivity in homoepitaxial HBDD films arises from electron correlation induced by critical boron doping concentration, rather than macroscopic grain boundaries. Prior studies of nano- and polycrystalline HBDD thin films, where the structural granularity dominates, have reported some form of anisotropy in the superconducting phase~\cite{Zhang_PhysRevApplied.12.064042,Sobaszek_Carbon.119337,Bhattacharyya_Crystals.12.1031, Bustarret_PTRSA.366.267}. The observed anisotropy was weaker in those HBDD samples, possibly due to boron concentration exceeding 10\textsuperscript{21}cm\textsuperscript{-3} and thus a stronger electron-phonon coupling as most of the impurity states are fully delocalized, hence masking any contribution from electron correlation~\cite{Okazaki_APL.106.052601, Yokoya_Nat2005.438.7068, Bustarret_PTRSA.366.267, Klein_PhysRevB.75.165313}. By studying samples in the critical doping regime and also eliminating structural granularity, we isolate electronic granularity, revealing three tunable phases with distinct symmetries of an intrinsic electronic order in this disordered system. If the granular superconductivity in single crystal diamond is a result of random boron incorporation in the lattice, the electrical transport should be isotropic. Hence, the emergent anisotropic order we observe in this critically disordered system must arise due to the competition between electron correlation and electron-phonon interactions. Figure~\ref{Fig EM3} visualizes this as a network of superconducting puddles (blue), fermionic regions (red), and incoherent pre-formed Cooper pairs (green). Without structural pinning; temperature and electromagnetic fields can tune the size and shape of these channels, leading to the three observed phase symmetries (Fig.~\ref{Fig 3}). Given the symmetry of the (100) plane, the source of the broken symmetry is currently known. However, as we show in the Supplementary Material~\cite{Supplementary_Materials}, there is a clear spatial anisotropy in the electrical transport that is consistent with the schematic picture shown in Fig.~\ref{Fig EM3}. A preview of such an ``unpinned" anisotropic order parameter manifesting as a distorted vortex lattice has been seen in scanning tunneling spectroscopy measurements~\cite{Sacepe_PhysRevLett.96.097006}. Fits to the Werthamer-Helfand-Hohenberg (WHH) model for the transport data also show deviation from the expected BCS exponents (Fig. S5 in the Supplementary Material ~\cite{Supplementary_Materials}). This also points to the validity of the IBRVB model where the phase transition from a dirty Mott metal to a superconductor near the critical doping concentration involves strong electron correlations with spontaneous formation of B\textsuperscript{+} and B\textsuperscript{-} fermionic channels~\cite{Baskaran_JSNM.21.45}. The MR measurements give us further information about the symmetries of these electron-electron interactions. 

Our observation of strongly anisotropic intrinsic granular superconductivity in critically doped homoepitaxial HBDD films highlights the role of electron correlations in the superconducting phase of samples near the metal-insulator transition, clearly indicating that boron doping in diamond induces electronic inhomogeneity in the superconducting phase, even in highly ordered, crystalline HBDD. The anisotropic order emerging within this electronic disorder can be tuned by temperature and magnetic field. Probing this anisotropic order using techniques other than transport, such as scanning tunneling microscopy, is crucial for gaining a deeper understanding of granular superconductivity in HBDD; this will provide more generalized insights into the superconducting phase of other doped group IV semiconductors as well. Controlling the emergent anisotropy could provide a route to higher critical temperatures in HBDD, as suggested by theoretical work on doping-induced disorder~\cite{Shirakawa_JPSJ.76.014711}. A recent study has shown tantalizing evidence that the possibly ordered incorporation of boron dopants into the diamond lattice and accompanying Bose-Einstein condensation may be responsible for qualitatively similar anomalous transport in polycrystalline HBDD~\cite{Sobaszek_Carbon.119337}. Finally, a complete understanding of the interplay between electron correlations and dopant-induced disorder in HBDD is relevant for applications in quantum technologies. For example, the recent demonstration of uniformly distributed NV centers in the vicinity of a HBDD layer~\cite{Lin_AdvFuncMat} allows us to envision exploiting electron correlations to facilitate information transfer across NV center qubits by using a magnetic field to control whether they interact with fermionic or bosonic patches. 
    
\section*{Acknowledgements} J.D. and N.S. thank Prof. Katja Nowack and Austin Kaczmarek (Department of Physics, Cornell University) for helpful discussions. J.D. also acknowledges Matthew Krebs (Department of Physics, The Pennsylvania State University) and Pramudit Tripathi (Department of Chemical Engineering, The Pennsylvania State University) for their assistance in the data analysis code. K.D.H., G.A.V.-L., and D.R.H. acknowledge support through startup funds from the Penn State Eberly College of Science, Department of Chemistry, College of Earth and Mineral Sciences, Department of Materials Science and Engineering, and Materials Research Institute. The authors also acknowledge the use of the Penn State Materials Characterization Lab. Primary support for this research was provided by U.S. Department of Energy Office of Science National Quantum Information Science Research Centers (Q-NEXT). We also acknowledge a seed grant from the Penn State Materials Research Institute. This work uses low-temperature transport facilities (DOI: 10.60551/rxfx-9h58) provided by the Penn State Materials Research Science and Engineering Center under award NSF-DMR 2011839.
\section*{Author Contributions} The project was designed by J.D., N.S., and D.D.A., with input from N.D. and J. H. J.D. synthesized and characterized the films, fabricated devices, and carried out electrical transport measurements with assistance from J. M., S. I. and A.R. and under the supervision of N.S., D.S., and L.L. K.D.H. and G.A.V.-L. performed FIB and TEM measurements under the supervision of D.R.H. J.D. and N.S. wrote the manuscript with feedback from all authors.\\
\section*{Data and Materials Availability} All data needed to evaluate the conclusions in the paper are present in the paper and/or the Supplementary materials.
\clearpage
\newpage
\section*{End Matter}
\begin{figure}
    \centering
    \includegraphics{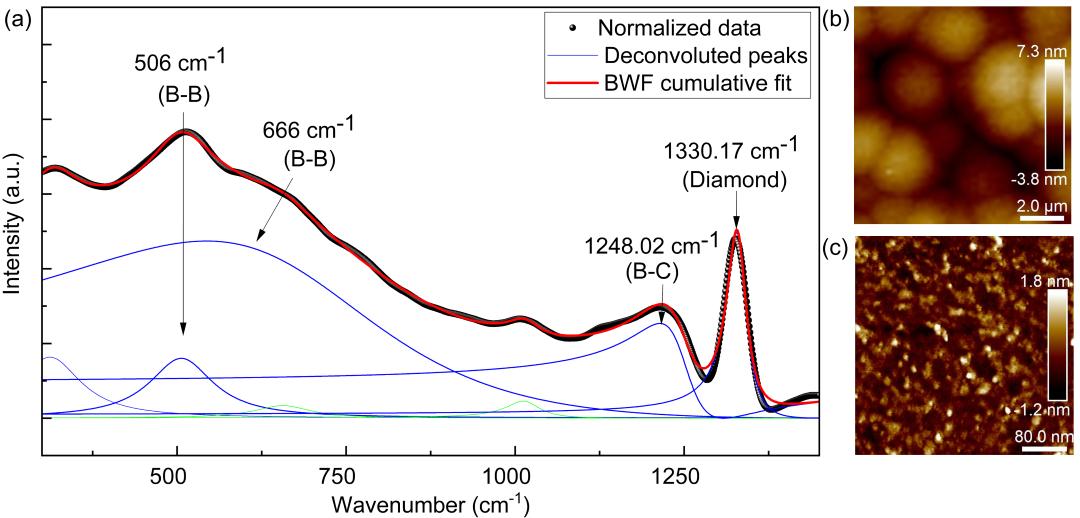}
    \caption{(a) Raman spectrum shows the modes expected of B-doping. (b) AFM scans of a 100 \textmu m\textsuperscript{2} area show very shallow island-like features with $Rq = 2.32$ nm. (c) Zoomed-in topography of a single island with $Rq = 0.5$ nm.
}
    \label{Fig EM1}
\end{figure}

To gain insights into the B incorporation, we use spatially resolved Raman spectroscopy (Fig. S1(b) in \cite{Supplementary_Materials}) \cite{Mortet_Carbon.115.279}. Previous analysis of Raman spectra of HBDD samples has shown that B incorporates in the diamond lattice either as a single substitutional B defect or as B-B dimers. Increased doping leads to more complex B incorporation forming BB-C-BB tetramers \cite{Sobaszek_Carbon.119337}. Figure \ref{Fig EM1}(a) shows the Raman spectrum measured in an HBDD thin film; a detailed analysis of this spectrum using the Breit-Wigner-Fano function identifies the different Raman modes \cite{Fano_PhysRev.124.1866} (Fig. S1(a) in \cite{Supplementary_Materials}). An empirical linear fit \cite{Mortet_Carbon.168.319} correlates the B-doping concentration to the Raman peak shift, providing an estimated B concentration of $5 \pm 0.4 \times 10^{20}$ cm\textsuperscript{-3}. This places our sample at the critical doping concentration limit for observing superconductivity in HBDD \cite{Okazaki_APL.106.052601,Winzer_PhysicaC.432.65}.
Additional characterization of the HBDD films is carried out using AFM measurements (Fig.~\ref{Fig EM1}(b),(c)). In thicker HBDD films ($t \geq 10$ \textmu m), we observe rougher surfaces with average root mean square roughness $Rq > 2.3$ nm, with pyramidal defects and non-uniform doping in spatial Raman maps (Fig. S1(c) in \cite{Supplementary_Materials}). This suggests that B doping incorporation variability can be a function of dominant diamond crystal face during synthesis and can be inferred from surface roughness and spatial Raman mapping.
\begin{figure}
    \centering
    \includegraphics{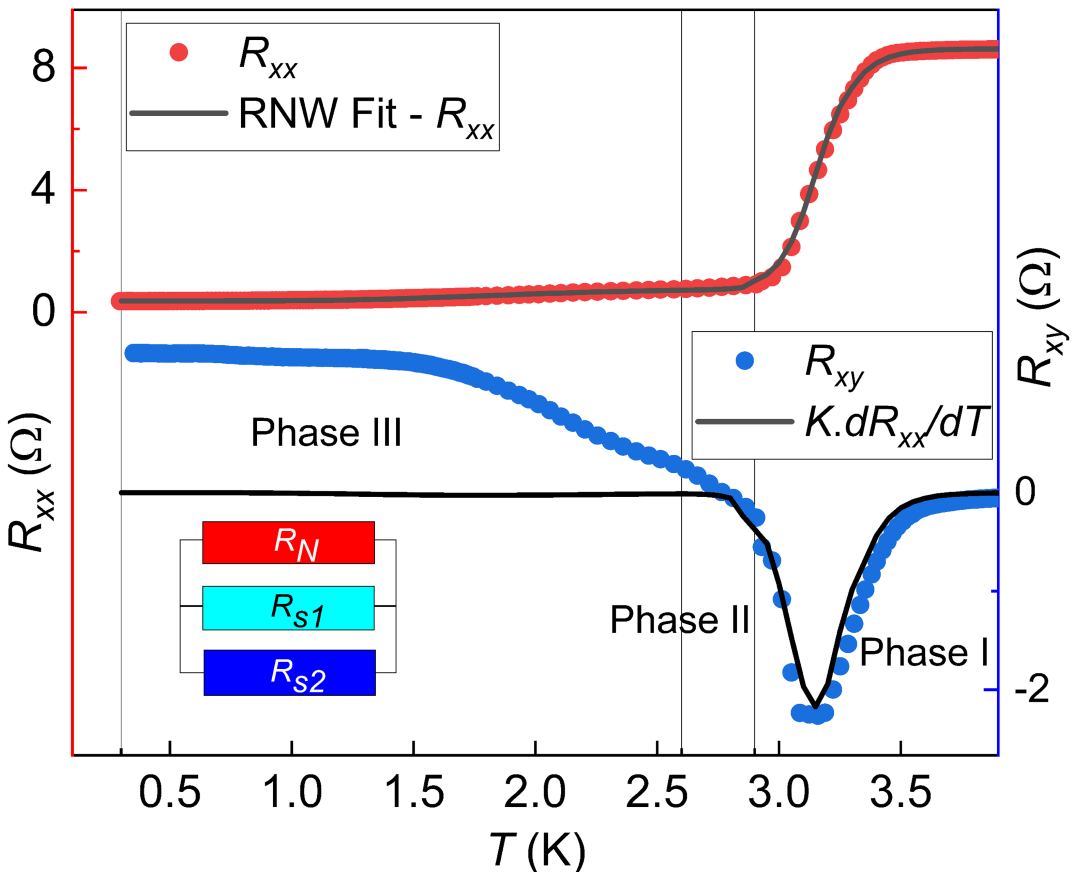}
    \caption{Resistor network model fits to \Rxx~vs. $T$ (red circles) and \Rxy~vs. $T$ (blue circles). The latter is fit to $Kd$\Rxx$/dT$.}
    \label{Fig EM2}
\end{figure}
AFM measurements for sample AE-1 (0.5 \textmu m thick) (Fig.~\ref{Fig EM1}(b)) shows a surface roughness of $Rq\sim 2.3$ nm with island-like features of an average diameter of about 2 $\mu$m. The island surface itself is smooth with an rms roughness of $Rq \sim 0.5$ nm (Fig.~\ref{Fig EM1}(c)) - an indirect indication of uniform boron doping. Additional TEM analysis shows diffraction patterns with (100) orientation across a greater than 9 \textmu m wide region, establishing these films as single crystalline (Fig. S2 in \cite{Supplementary_Materials}).
\begin{figure}
    \centering
    \includegraphics{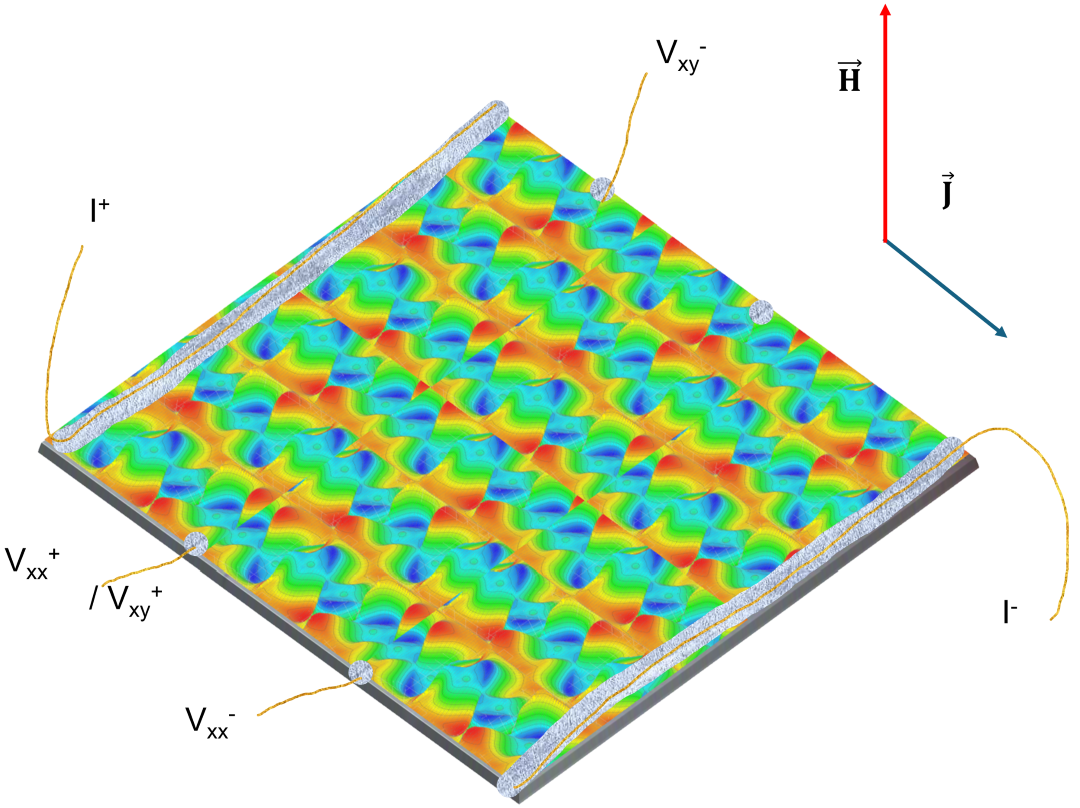}
    \caption{Schematic illustration of electronic granularity in a HBDD sample. Red regions represent fermionic channels while blue regions represent bosonic puddles. Contact configuration for the resistivity measurements is denoted by the silver contacts and gold wire. The edges of this sample are in the $\langle 110 \rangle $ family. }
    \label{Fig EM3}
\end{figure}
\begin{figure}
    \centering
    \includegraphics{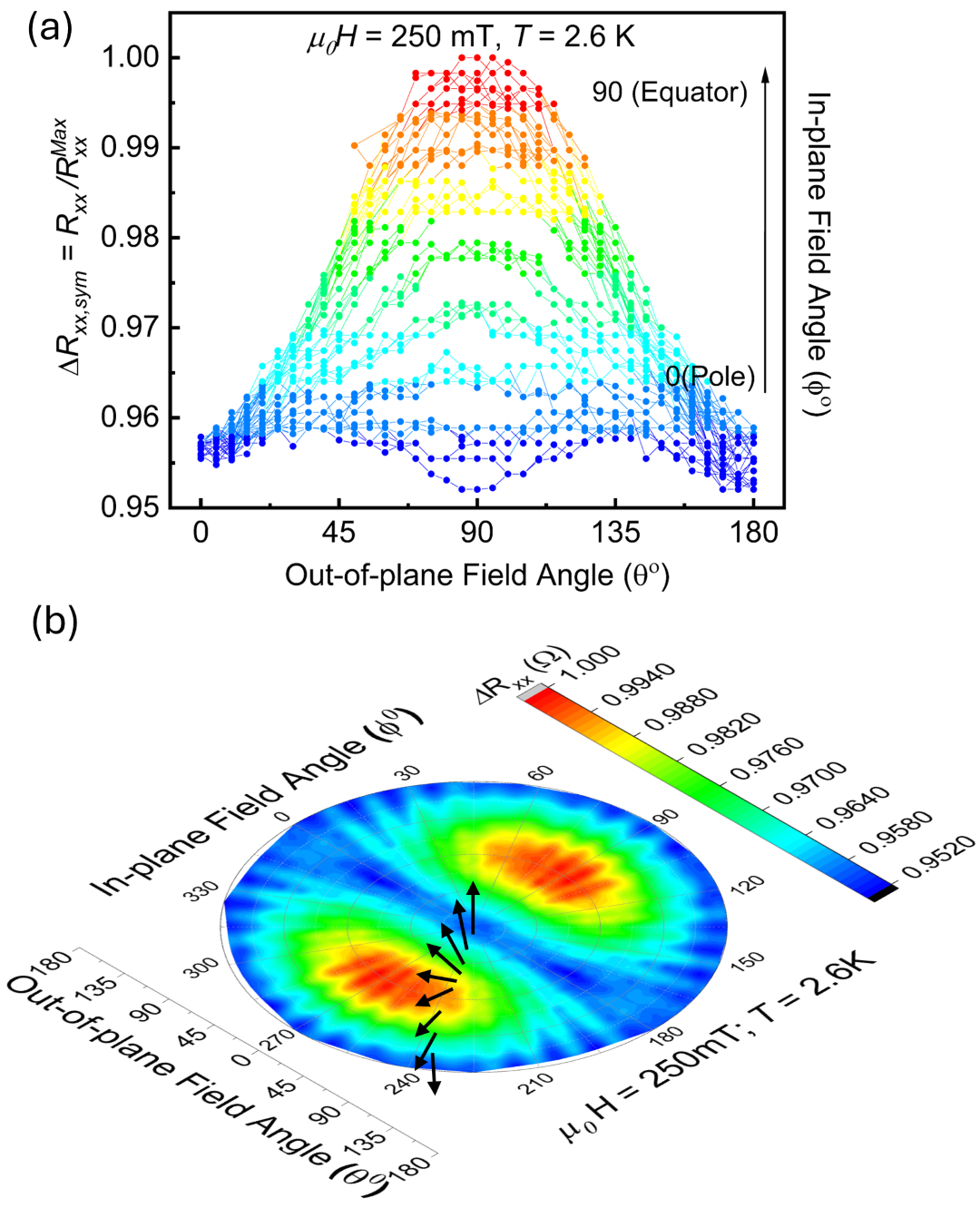}
    \caption{(a) Representative plot showing the variation of $R_{xx,sym}$ as $\vec{H}$ is rotated from OP to IP through angle $\theta$ at $T = 2.62$ K and with $\mu_0 |\vec{H}|=250$ mT. Curves from bottom to top correspond to increasing IP field angle ($\phi$). (b) Azimuthal equidistant projection of the data in (a) showing angular anisotropy in the MR which is dependent on $\vec{H} \cdot \vec{J}$ .
}
    \label{Fig EM4}
\end{figure}
We fit the temperature dependence of \Rxx~using one fermionic resistor ($R_N$) and two bosonic resistors ($R_{S1}$ and $R_{S2}$) as shown in Fig.~\ref{Fig EM2} (\cite{Supplementary_Materials}). Near \Tco, the fermionic channel dominates (Phase I). Around $T_c^{50\%}$, the bosonic channel starts to dominate because it offers lower resistance (Phase II). Finally, below \Tcoff, the bosonic channels reach their maximum possible coherence but residual fermionic channels lead to low non-zero resistance (Phase III). A similar resistor network approach can be used to fit the Hall anomaly data as well, shown on the right axis of Fig.~\ref{Fig EM2}~\cite{Segal_PRB.83.094531}. The model fits Phases I and II of our data very well but the deviations in Phase III suggest the need of a denser resistor network.
Figure~\ref{Fig EM4}(a) shows the \Rxxsym~variation while rotating $\vec{H}$ from OP to in-plane (IP). The x-axis represents the angle $\theta$ between $\vec{H}$ and $\vec{z}$. Different curves from bottom to top correspond to an increase in azimuthal angle $\phi$ between the projection of $\vec{H}$ on the $x-y$~plane ($\vec{H}_{xy}$) and the IP-current  $\vec{J}$ along the [100] direction. Figure~\ref{Fig EM4}(b) is the azimuthal equidistant projection of the data in Fig.~\ref{Fig EM4}(a). At temperatures below \Tco, resistance is highly sensitive to the magnetic field direction and is symmetric with respect to the current direction. High resistance (red) regions lying in a 45$^{\circ}$~conical section around the equatorial plane correspond to higher IP magnetic field components. Low resistance (blue) regions are confined around the poles and correspond to the OP field components. The color intensity is proportional to the magnitude of the angular MR variation within each plot. The rotation of the field vector has been marked with arrows for $\phi=240^{\circ}$. 

%

\end{document}